\documentclass[11pt,twoside]{article}

\usepackage{asp2006}
\usepackage{epsf}

\begin{document}

\title{Shell Galaxies: Dynamical Friction, Gradual Satellite Decay and Merger Dating}   

\author{
Ivana Ebrov\'a,$^{1,2}$
Bruno Jungwiert,$^2$
Gabriela Canalizo,$^3$
Nicola Bennert,$^4$
and Lucie J\'ilkov\'a$^5$
}

\affil{
$^1$Faculty of Mathematics and Physics, Charles University in Prague, Ke~Karlovu~3, CZ-121~16 Prague, Czech Republic}
\affil{$^2$Astronomical Institute, Academy of Sciences of the Czech Republic, Bo\v{c}n\'{i} II 1401/1a, CZ-141 31 Prague, Czech Republic}
\affil{$^3$Institute of Geophysics and Planetary Physics \& Dept. of Physics, University of California, Riverside, CA 92521, USA}
\affil{$^4$Dept. of Physics, Univ. of California, Santa Barbara, CA 93106, USA}
\affil{$^5$Dept. of Theoretical Physic and Astronomy, Faculty of Science,
Masaryk University, Kotl\'a\v rsk\'a 2, CZ-611 37 Brno, Czech Republic}


\begin{abstract}
With the goal to refine modelling of shell galaxies and the use of shells to probe the merger history, we develop a new method for implementing dynamical friction in test-particle simulations of radial minor mergers. The friction is combined with a gradual decay of the dwarf galaxy. The coupling of both effects can considerably redistribute positions and luminosities of shells; neglecting them can lead to significant errors in attempts to date the merger.
\end{abstract}

\section{Shells as Probes of the Host Galaxy Merger History}
Shell galaxies contain faint arc-like stellar features. It is widely believed that shells are a signature of a merger experienced by the host galaxy. 
They contain at most a few percent of the overall galaxy luminosity, and their contrast is usually very low.
The model of a radial merger of a giant elliptical with a smaller galaxy
(a spiral or a dwarf elliptical)
(\citealp{quinn84}; \citealp{dupraz86}; \citealp{hernquist88}) 
seems to be the most successful in reproducing regular shell systems. When a small galaxy enters the sphere of influence of a giant elliptical on a close-to-radial trajectory, it disintegrates and its stars begin to oscillate in the potential of the giant. At their turning points, where the stars tend to spend most of their time, they pile up and produce arc-like enhancements in the luminosity profile of the host galaxy.

\begin{figure}
\plotone{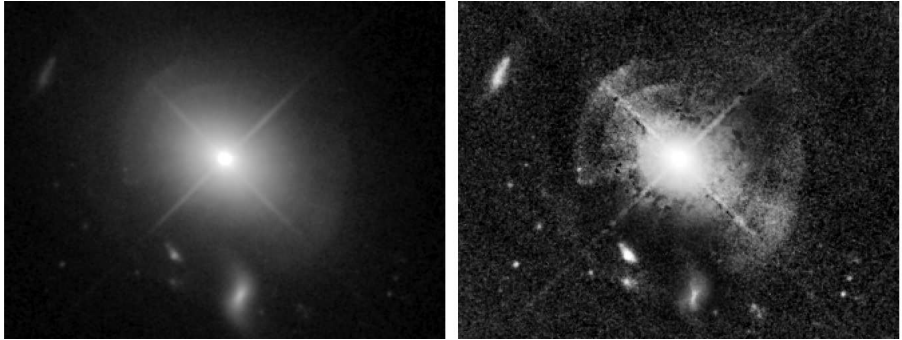}
\caption{
Deep HST/ACS images of the host galaxy of the quasar MC2 1635+119, so far the only known shell galaxy with a quasar (\citealp{canalizo07}; \citealp{bennert08}). The left panel shows the original image, the right one the residual after the subtraction of the fitted smooth light profile.
}\label{mc2}
\end{figure}

Attempts to date a merger from observed positions of shells have been made in previous works. Recently, \citet{canalizo07} presented HST/ACS observations of spectacular shells in a quasar host galaxy (Fig.\,\ref{mc2}) and, by simulating the position of the outermost shell by means of restricted N-body simulations, 
attempted to put constraints on the age of the merger. They concluded that it occurred a few hundred Myr to $\sim$~2~Gyr ago, 
supporting a potential causal connection between the merger, the post-starburst ages in nuclear stellar populations, and the quasar.
A typical delay of 1--2.5~Gyr between a merger and the onset of quasar activity is suggested by both N-body simulations \citep{springel05} and observations \citep{ryan08}. 
It might therefore appear reassuring to find a similar time lag between the merger event and the quasar ignition in a study of an individual spectacular object. However, caution must be exercised in estimating merger ages from the location of shells (see below).     

\section{Dynamical Friction and Gradual Decay of the Satellite}

\begin{figure}[!ht]
\plotone{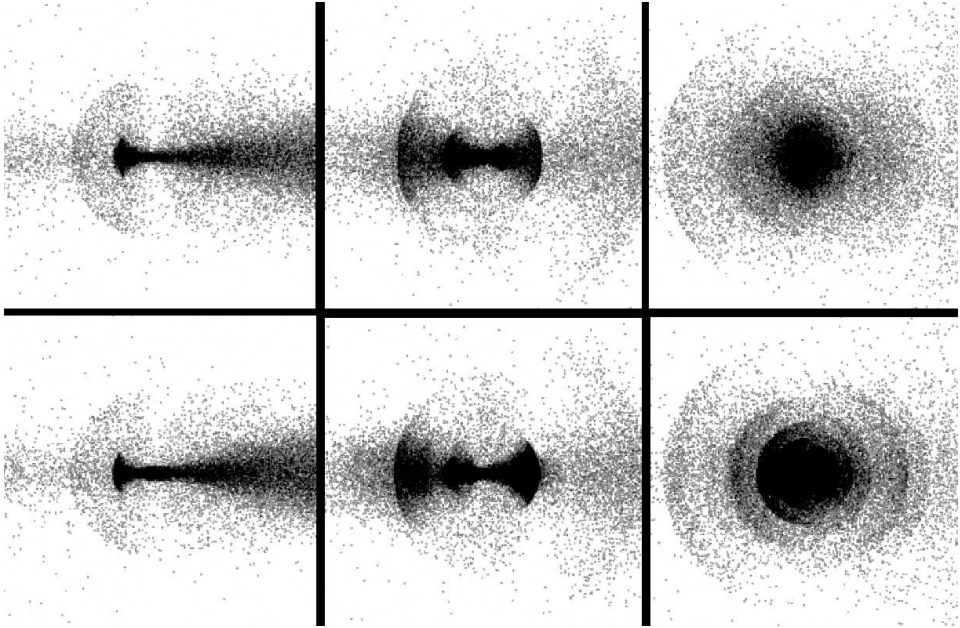}
\caption{
Three snapshots of simulations (3.5, 5 and 7~Gyr after the first passage of 
the satellite, coming from the right, through the center of the primary)
without (upper row) and with (bottom row) dynamical friction and gradual 
disruption (in the first case, the dwarf instantly disrupts during the first passage). Only stars of the dwarf are shown. Each box, centered on the primary, 
shows 300$\times$300~kpc.
}\label{film}
\end{figure}

While the shell formation, once the dwarf galaxy is disrupted, is basically a test-particle phenomenon, the gradual decay of the satellite as well as 
its braking by dynamical friction against the primary can considerably affect the energy distribution of oscillating stars, and thus the positions
and the brightness of shells. The dynamical friction effect was first pointed out by \citet{dupraz87} and also discussed by \citet{hernquist88}, while the
gradual decay, with friction neglected, was modelled by \citet{heisler90}. However, coupling of these phenomena was never modelled in much detail.
Our goal is to improve restricted N-body simulations of shells created in minor mergers by a)~inclu\-ding dynamical friction, b)~improving its implementation by avoiding the use of the Chandrasekhar formula, c)~coupling it to the gradual decay, d)~taking into account the present state of knowledge of stellar and dark matter distributions in both 
giant and dwarf ellipticals. A detailed description is beyond the scope of this paper. Here, we confine ourselves to a simple example of a radial minor merger (Fig.\,\ref{film}), instructive in showing how an observed shell structure could be misinterpreted in terms of the merger time scale
(and of the relative pre-merger motion) if dynamical friction and gradual decay were neglected.

In test-particle simulations, the Chandrasekhar formula is commonly 
used to include dynamical friction. Its relative simplicity is made possible, among others, by the oversimplifying assumption of homogeneity of the stellar and dark matter distributions. To avoid it, we used the axial symmetry of our merger configuration to simplify the integrals over impact parameters and velocity distributions so that they can be solved numerically. 
The mass of the satellite, a key quantity
for the efficiency of dynamical friction, is gradually lowered in proportion to the mass located beyond its evolving tidal radius.

\begin{figure}[!ht]
\plotone{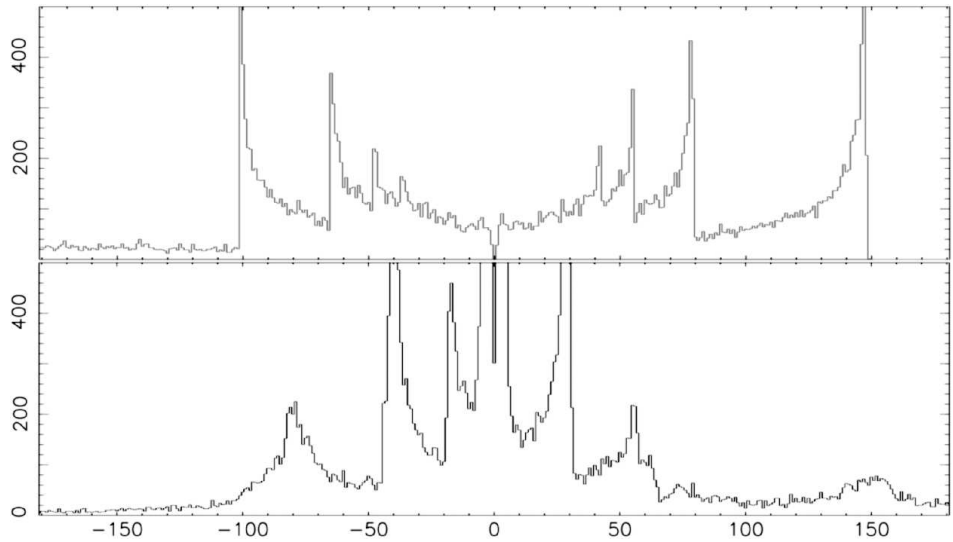}
\caption{
Histograms of galactocentric distances of stars (in kpc) originally 
belonging to the dwarf, at 5~Gyr. {\it Top:} Instantaneous 
disruption, no friction; {\it Bottom:} Gradual disruption plus 
friction. Distances are measured from the center of the primary, and 
plotted separately for positions on the side from which the 
satellite came and those on the opposite one (plus/minus sign). 
}\label{dm}
\end{figure}

The introduction of dynamical friction and gradual decay dramatically changes the appearance of shells as can be seen in histograms of particles' galactocentric distances (Fig.\,\ref{dm}, corresponding to central snapshots
of Fig.\,\ref{film}).
While the position of the outermost shell is not much affected, its brightness is drastically lowered. The other shells are shifted and new generations of shells are added during each successive passage 
of the dwarf. Easily inferring the age of the collision is rendered impossible
(as already pointed out by \citealp{dupraz87}).
The shell systems in Fig.\,\ref{dm}, 
both having the outermost shell at $+$150~kpc, are seen 5~Gyr 
after the first passage of the two galaxies through each other. 
If we observationally identified the leftmost shell (at $-$80~kpc in Fig.\,\ref{dm}, lower panel) as being the outermost one, we would mistakenly estimate the merger age to be only $\sim$~2.5~Gyr. We would also wrongly
determine the direction from which the dwarf came: assuming the classical picture (based on simulations without friction and with instantaneous disruption), the outermost shell would be located on the side from which the satellite came, so we would conclude it went from the left while the opposite is true.

\section{Conclusions}

Using even the outermost observed shell to date a merger, and basing
on it a support for a causal connection between the merger and the quasar, 
is very uncertain. Supposedly, the first formed shell (observed as 
the outermost one if still undissolved and bright enough) 
is the least affected by dynamical friction
(since it is formed out of stars released during the first satellite's 
passage) and thus the most reliable for merger dating. 
In our example, this first shell is very weak due
to the gradual decay of the satellite. If missed in observations, 
the merger age would be underestimated by $\sim$~2.5~Gyr; in reality, 
it is twice as old. 

\acknowledgements This project is supported by the Inst. Research Plan AV0Z10030501 of the Academy of Sciences of the Czech Republic, and by the grants 205/08/H005 of the Czech Science Foundation and LC06014 (Center for Theoretical Astrophysics) of the Czech Ministry of Education.


\end{document}